\documentclass{revtex4}

\usepackage[usenames,dvipsnames]{color}
\usepackage{amsfonts,amssymb,amsmath,amsthm}
\usepackage{bm}
\usepackage[a4paper]{geometry}
\usepackage{pdflscape}
\usepackage{graphicx}
\usepackage{mathrsfs}

\definecolor{darkred}{rgb}{.8,0,0}

\definecolor{darkblue}{rgb}{0,0,.7}

\newcommand{\eps}{\varepsilon}

\newcommand{\rev}[1]{\widetilde{#1}}
\newcommand{\blade}[2]{{#1}_1 \wedge \ldots \wedge {#1}_{#2}}

\newcommand{\uline}[1]{\underline{#1}}
\newcommand{\oline}[1]{\overline{#1}}

\theoremstyle{plain}
\newtheorem*{TrPhysMot}{Transformation of physical motions}
\newtheorem*{SymHam}{Symmetry transformation}
\newtheorem*{Noether}{Noether theorem}
\newtheorem*{GenSymHam}{Generalized symmetry transformation}
\newtheorem*{GenNoether}{Generalized Noether theorem}

\begin{document}

\title{Classical field theories from Hamiltonian constraint: \\
Symmetries and conservation laws}

\author{V\'{a}clav Zatloukal}

\email{zatlovac@fjfi.cvut.cz}

\homepage{http://www.zatlovac.eu}

\affiliation{\vspace{3mm}
Faculty of Nuclear Sciences and Physical Engineering, Czech Technical University in Prague, \\
B\v{r}ehov\'{a} 7, 115 19 Praha 1, Czech Republic \\
}

\affiliation{
Max Planck Institute for the History of Science, Boltzmannstrasse 22, 14195 Berlin, Germany
}

\begin{abstract}
We discuss the relation between symmetries and conservation laws in the realm of classical field theories based on the Hamiltonian constraint. In this approach, spacetime positions and field values are treated on equal footing, and a generalized multivector-valued momentum is introduced. We derive a field-theoretic Hamiltonian version of the Noether theorem, and identify generalized Noether currents with the momentum contracted with symmetry-generating vector fields. Their relation to the traditional vectorial Noether currents is then established.

Throughout, we employ the mathematical language of geometric algebra and calculus.
\end{abstract}

\maketitle

%%%%%%%%%%%%%%%%%%%%%%%%%%%%%%%%%%%%%%
\section{Introduction}
%%%%%%%%%%%%%%%%%%%%%%%%%%%%%%%%%%%%%%

In Chapter 3 of monograph \cite{RovelliQG}, it has been explained how classical field theory can be formulated in a way that treats spacetime points $x$ and field values $\phi$ in a symmetric manner. These are collectively called \emph{partial observables}, and their possible outcomes are represented by points $q=(x,\phi)$ in a $D+N$-dimensional \emph{configuration space} $\mathcal{C}$. (In this article, for simplicity, we will have in mind a Euclidean configuration space, although an extension of the theory to pseudo-Euclidean spaces should be straightforward.) Collections $\gamma$ of points $q$ that form $D$-dimensional surfaces in $\mathcal{C}$ are called \emph{motions}. They represent correlations between the partial observables, i.e., gather possible outcomes of their simultaneous measurements. Note that in the traditional approach to field theory, $\gamma$ would be regarded as an $N$-component function $\phi(x)$.

Classical field theory predicts the class of motions that can occur in nature --- the \emph{physical (or classical) motions} $\gamma_{\rm cl}$. These are obtained as extremals of a variational principle (see \cite[Ch.~3.3]{RovelliQG}), which we phrase in the geometric algebra language as follows (see our previous article \cite{ZatlCanEOM} for details):

A surface $\gamma_{\rm cl}$ with boundary $\partial \gamma_{\rm cl}$ is a physical motion, if the couple $(\gamma_{\rm cl},P_{\rm cl})$ extremizes the (action) functional
\begin{equation} \label{Action}
\mathcal{A}[\gamma,P] = \int_\gamma P(q) \cdot d\Gamma(q)
\end{equation}
in the class of pairs $(\gamma,P)$, for which $\partial\gamma = \partial\gamma_{\rm cl}$ and the $D$-vector-valued momentum $P$ defined along $\gamma$ satisfies
\begin{equation} \label{HamConstraint}
H(q,P(q)) = 0 ~~~~~ \forall q \in \gamma .
\end{equation}

The surface $\gamma$, with its infinitesimal oriented element $d\Gamma$, and the momentum field $P$ are varied independently, as long as they satisfy the \emph{Hamiltonian constraint} (\ref{HamConstraint}). We will assume, for simplicity, that the function $H$ (the \emph{relativistic Hamiltonian}, or simply \emph{Hamiltonian}) is scalar-valued. (A theory with multiple constraints can be derived in full analogy.)

In Ref.~\cite{ZatlCanEOM}, we used the method of Lagrange multipliers to promote the Hamiltonian constraint into the action,
\begin{equation} \label{ActionAugm}
\mathcal{A}[\gamma,P,\lambda] 
= \int_\gamma \left[ P \cdot d\Gamma - \lambda H(q,P) \right] ,
\end{equation}
in order to derive the following \emph{canonical equations of motion}:
\begin{subequations} \label{CanEOM}
\begin{align} 
\label{CanEOM1}
\lambda \, \partial_P H(q,P) &= d\Gamma , 
\\ \label{CanEOM2}
(-1)^D \lambda \, \dot{\partial}_q H(\dot{q},P) 
&= \begin{cases}
d\Gamma \cdot \partial_q P & ~~{\rm for}~ D=1 \\
(d\Gamma \cdot \partial_q) \cdot P &  ~~{\rm for}~ D>1 ,
\end{cases}
\\
\label{CanEOM3} 
H(q,P) &= 0 .
\end{align}
\end{subequations}
(The overdot is used to specify the scope of differentiation.)
For $D=1$, these equations describe relativistic or non-relativistic particles (depending on the definition of $H$), while for $D>1$ they provide a first-order formulation of, for example, string or scalar field theory (see examples in \cite{ZatlCanEOM}).

%Moreover, we discussed the Hamilton-Jacobi theory, and proposed a field-theoretic local Hamilton-Jacobi equation
%\begin{equation} \label{HJeq}
%H(q,\partial_q \wedge S) = 0 ,
%\end{equation}
%where $S(q)$ is a $(D-1)$-vector field on $\mathcal{C}$.

The aim of this article is to find an appropriate generalization of the Noether theorem \cite{Noether1918}, which relates symmetries and conservation laws, within the framework of the above-outlined Hamiltonian field theory. For this purpose, in Sec.~\ref{sec:Sym}, we study transformations of the configuration space $\mathcal{C}$, and specify conditions, namely, Eq.~(\ref{SymCrit}), under which they represent \emph{symmetries} of the physical system. In Sec.~\ref{sec:ConsLawsFromSym}, we combine infinitesimal symmetries with the canonical equations of motion to derive the conservation law (\ref{ConsLaw}), and hence the Hamiltonian Noether theorem. The respective conserved quantity is a $D-1$-vector-valued Noether current  obtained by contracting the momentum with a symmetry-generating vector field. In fact, more general conservation laws can be obtained by considering transformations that preserve the action only up to a boundary term. This is achieved in Sec.~\ref{sec:GenSym}, Eq.~(\ref{GenConsLaw}). 
%At the end of this section we also briefly discuss a field-theoretic generalization of the Poisson brackets.

The general theory is illustrated with a number of examples in Sec.~\ref{sec:Examples}. First, we discuss symmetries of non-relativistic mechanical systems in the sense of Sec.~\ref{sec:Sym}, as well as in the generalized sense of Sec.~\ref{sec:GenSym}. The latter are exemplified by the Galilei transformations. The second example is concerned with an $N$-component scalar field theory, or, more generally, with the De~Donder-Weyl Hamiltonian field theory. We make contact with the standard treatment of scalar fields as functions on spacetime, and associate our multivector-valued Noether currents with their traditional vectorial counterparts. In particular, we examine spacetime translations and rotations, and rotations in the field space, and verify that they yield conservation of the energy-momentum and angular-momentum tensors, and internal vectorial currents, respectively. In the last example, we find symmetries of a bosonic string theory to be arbitrary rigid rotations and translations.

Let us recall that all manipulations are performed in the mathematical formalism of geometric (or Clifford) algebra and calculus developed by D. Hestenes \cite{Hestenes} (see also Ref.~\cite{DoranLas}). This coordinate-free language unifies several areas of mathematics, such as differential forms, spinors, complex analysis, and others. The reader can find a brief, but concise, introduction into this formalism in the appendix of our previous article \cite{ZatlCanEOM}. In Appendix~\ref{sec:GAGC} of the present article, we recall some basic notions, and discuss, in particular, transformations and induced mappings, which play an essential role in what follows. A subsection about the geometric algebra approach to rotations is also included.

%%%%%%%%%%%%%%%%%%%%%%%%%%%%%%%%%%%%%%
\section{Symmetries in the Hamiltonian approach}
%%%%%%%%%%%%%%%%%%%%%%%%%%%%%%%%%%%%%%
\label{sec:Sym}

\begin{figure} 
\includegraphics[scale=1]{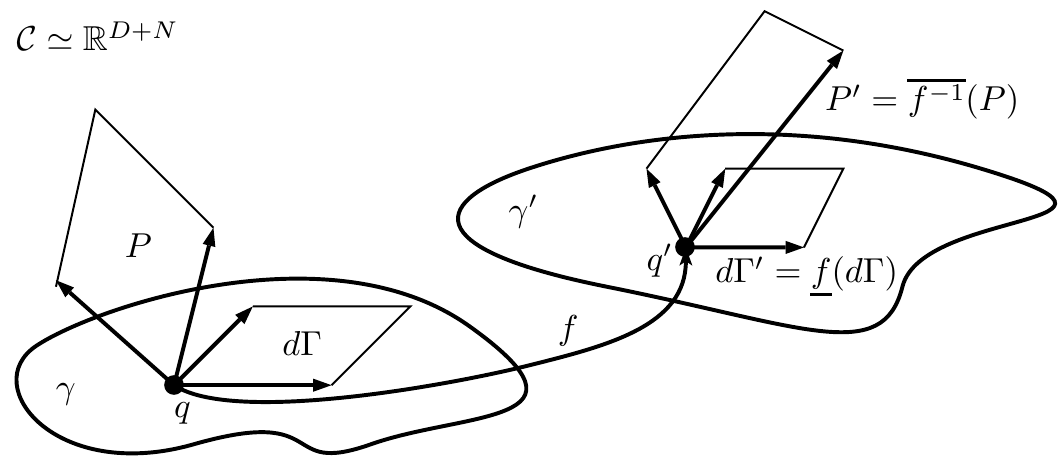}
\caption{Transformation of motions, surface elements, and momenta under a diffeomorphism $f$.}
\label{fig:Transf}
\end{figure}

A transformation of the configuration space $\mathcal{C}$ of partial observables is expressed mathematically by a diffeomorphism $f:\mathcal{C}\rightarrow\mathcal{C}$ (see Fig.~\ref{fig:Transf}), 
which maps a generic surface $\gamma$ to another surface 
\begin{equation}
\gamma'=\{q'=f(q)\,|\,q \in \gamma \} ,
\end{equation}
while the surface elements on $\gamma$ and $\gamma'$ are related by the induced outermorphism $\underline{f}$, 
\begin{equation}
d\Gamma'(q') = \underline{f}(d\Gamma(q);q) .
\end{equation}
(Transformations and induced mappings are discussed in Appendix \ref{sec:TrGC}.)

Note that $f$ is an \emph{active} transformation --- it is a mapping between points of the configuration space $\mathcal{C}$. In the dual picture, one could consider passive transformations, i.e., changes of coordinates on $\mathcal{C}$. Since we are working completely without coordinates, all transformations in this article are viewed as active.

A relation between the momentum fields on $\gamma$ and $\gamma'$ is established by demanding that the inner product $P \cdot d\Gamma$, and hence the action (\ref{Action}), be invariant under $f$. This is achieved by postulating the transformation rule
\begin{equation} \label{TrP}
P'(q') = \overline{f^{-1}}(P(q);q) .
\end{equation}
Invariance of the action then implies the following:
\begin{TrPhysMot}
Consider an arbitrary diffeomorphism $f:\mathcal{C}\rightarrow\mathcal{C}$. If $\gamma_{\rm cl}$ is a physical motion of a system with Hamiltonian $H$, then
\begin{equation} \label{TrPhysMotion}
\gamma'_{\rm cl} = \{q'=f(q)\,|\,q \in \gamma_{\rm cl} \}
\end{equation}
is physical motion of a system with Hamiltonian $H'$, defined by
\begin{equation} \label{TrHam}
H'(q',P') = H(q,P) ,
\end{equation}
where $P'=\overline{f^{-1}}(P;q)$.
\end{TrPhysMot}
(An explicit proof of this claim on the level of canonical equations of motion is provided in Appendix \ref{sec:TrPhysMotCanEOM}.)

%Note that the definition of $P'$, Eq.~(\ref{TrP}), introduces an additional $q$-dependency into the Hamiltonian $H'$ as compared to $H$, due to the $q$-dependency of $\overline{f^{-1}}$.

We call a transformation $f$ a \emph{symmetry} if it maps physical motions to physical motions of the same physical system. This is the case when $H$ and $H'$ coincide, i.e., when
\begin{equation}
H'(q',P') = H(q',P') .
\end{equation}
As an immediate consequence of the definition (\ref{TrHam}) we then obtain:
\begin{SymHam}
A transformation $f$ is a symmetry of a physical system described by the Hamiltonian $H$ (or, in short, a symmetry of $H$), if
\begin{equation} \label{SymCrit}
H(f(q),\overline{f^{-1}}(P;q)) = H(q,P) .
\end{equation}
For infinitesimal transformations $f(q) = q + \eps v(q)$, $\eps \ll 1$, determined by a vector field $v$, Eq.~(\ref{SymCrit}) takes the form
\begin{equation} \label{SymCritInfsm}
v \cdot \dot{\partial}_q H(\dot{q},P)
- \big( \dot{\partial}_q \wedge (\dot{v} \cdot P) \big) \cdot \partial_P H(q,P)
= 0 .
\end{equation}
\end{SymHam}
Eq.~(\ref{SymCritInfsm}) is obtained from Eq.~(\ref{SymCrit}) in a straightforward way by utilizing Eq.~(\ref{GCinfsmInvTr}), i.e., the infinitesimal version of the transformation rule (\ref{TrP}).

In fact, the two Hamiltonians $H$ and $H'$ need not be identical for all $q$ and $P$. It suffices that $H(q,P)=0$ if and only if $H'(q,P)=0$, since then they still define, via the Hamiltonian constraint, the same dynamics.

Let us add a few words about infinitesimal transformations. They arise from one-parameter groups of transformations $f_\tau(q)$ in the small-$\tau$ limit, when we can approximate
\begin{equation}
f_\tau(q) \approx q + \tau v(q)
~~~,~~~ v(q) = \partial_\tau f_\tau(q)|_{\tau=0} .
\end{equation}
Conversely, to any vector field $v(q)$ corresponds a \emph{flow} $f_\tau(q)$, which can be regarded as a group of transformations parametrized by $\tau$. By means of the \emph{Lie series} \cite[Ch.~1.3]{Olver}, we can write an explicit formula
\begin{equation} \label{LieSeries}
f_\tau(q) = e^{\tau v \cdot \partial_q} q 
= q + \tau v + \frac{\tau^2}{2!} (v \cdot \partial_q) v + \ldots .
\end{equation}

%%%%%%%%%%%%%%%%%%%%%%%%%%%%%%%%%%%%%%
\section{Conservation laws from symmetries}
%%%%%%%%%%%%%%%%%%%%%%%%%%%%%%%%%%%%%%
\label{sec:ConsLawsFromSym}

The symmetries of a physical system are imprinted in its Hamiltonian function $H(q,P)$, and can be explored by analyzing Eqs.~(\ref{SymCrit}) or (\ref{SymCritInfsm}) without any reference to the equations of motion.

However, when the system is assumed to follow a classical trajectory, then the symmetries induce \emph{conservation laws}. This fact is derived almost instantly in the Hamiltonian constraint formalism. Substituting canonical equations (\ref{CanEOM1}) and (\ref{CanEOM2}), respectively, into the first and the second term in Eq.~(\ref{SymCritInfsm}), we find (for $D>1$)
\begin{equation}
(-1)^D v \cdot \big( (d\Gamma \cdot \partial_q) \cdot P \big)
- \big( \dot{\partial}_q \wedge (\dot{v} \cdot P) \big) \cdot d\Gamma
 = 0 ,
\end{equation}
which can be rearranged using basic geometric algebra identities, Eq.~(\ref{GAident0}),
\begin{equation}
(d\Gamma \cdot \dot{\partial}_q) \cdot (\dot{P} \cdot v) +
(d\Gamma \cdot \dot{\partial}_q) \cdot (P \cdot \dot{v})
= 0 ,
\end{equation}
and finally combined into a single term to yield the equation
\begin{equation}
(d\Gamma \cdot \partial_q) \cdot (P \cdot v)
= 0 .
\end{equation}
The derivation for the case $D=1$ is fully analogous. 

The above considerations are summarized by the following Hamiltonian version of the celebrated
\begin{Noether}
If $f(q)=q+\eps v(q)$ is an infinitesimal symmetry of $H$, i.e., if Eq.~(\ref{SymCritInfsm}) holds, then the solutions of the canonical equations of motion (\ref{CanEOM}) satisfy the conservation law 
\begin{align} \label{ConsLaw}
d\Gamma \cdot \partial_q \, (P \cdot v)
&= 0  \hspace{10mm}{\rm for}~ D=1  \nonumber\\
(d\Gamma \cdot \partial_q) \cdot (P \cdot v)
&= 0 \hspace{10mm}{\rm for}~ D>1 .
\end{align}
\end{Noether}

The quantities that obey conservation laws play a distinguished role in physics. The Noether theorem therefore grants a special status to the $D-1$-vector $P \cdot v$, and clearly displays the importance of the momentum multivector $P$ not only in particle mechanics, but also in classical field theory.

%%%%%%%%%%%%%%%%%%%%%%%%%%%%%%%%%%%%%%
\section{Generalized symmetries and conservation laws}
%%%%%%%%%%%%%%%%%%%%%%%%%%%%%%%%%%%%%%
\label{sec:GenSym}

In fact, the transformation $f$ does not necessarily have to satisfy Eq.~(\ref{SymCrit}) in order to map physical motions to physical motions of the same system. In this section we show that a more general condition, namely, Eq.~(\ref{GenSymCrit}) below, is sufficient to guarantee that $f$ be a symmetry, and we derive an accordingly generalized version of the Noether theorem.

We start with the observation that Hamiltonians $H(q,P)$ and
\begin{equation} \label{DefHF}
H_F(q,P) := H(q, P + \partial_q \wedge F) ,
\end{equation}
where $F(q)$ is an arbitrary $D-1$-vector-valued function, are equivalent in the sense that whenever $\gamma_{\rm cl}$ is a classical motion for the Hamiltonian $H$ with the corresponding classical momentum $P_{\rm cl}$, then $\gamma_{\rm cl}$ is also a classical motion for $H_F$ with shifted classical momentum $P_{\rm cl}-\partial_q\wedge F$. The proof of this claim follows directly from the canonical equations of motion, and the fact that $\partial_q \wedge \partial_q \wedge F = 0$. On the action level, we note that the extended actions, Eq.~(\ref{ActionAugm}), corresponding to $H$ and $H_F$, respectively, differ only by a boundary term:
\begin{equation}
\mathcal{A}_F[\gamma,P - \partial_q \wedge F,\lambda] 
= \mathcal{A}[\gamma,P,\lambda] 
- \int_{\partial \gamma} F \cdot d\Sigma ,
\end{equation}
where $\partial\gamma$ is the boundary of $\gamma$, and $d\Sigma$ its surface element.

Now, $\gamma_{\rm cl}$ is related by $f$ to a physical motion $\gamma'_{\rm cl}$ of the Hamiltonian $H'(q',P')=H_F(q,P)$ (recall Eqs.~(\ref{TrPhysMotion}) and (\ref{TrHam})). In order for $f$ to be a symmetry, we require that $H'$ coincide with the original Hamiltonian $H$, meaning
\begin{equation}
H(q',P') = H_F(q,P) .
\end{equation}
Taking into account the definitions (\ref{TrP}) and (\ref{DefHF}), we obtain the following proposition:
\begin{GenSymHam}
A transformation $f$ is a generalized symmetry of H, if there exists a $D-1$-vector-valued function $F(q)$ such that
\begin{equation} \label{GenSymCrit}
H(f(q),\overline{f^{-1}}(P;q)) = H(q,P + \partial_q \wedge F) .
\end{equation}
For infinitesimal transformations $f(q) = q + \eps\, v(q)$, $\eps \ll 1$, we may assume that $F \rightarrow 0$ as $\eps \rightarrow 0$, that is, $F = \eps W + O(\eps^2)$. Then, Eq.~(\ref{GenSymCrit}) takes the form
\begin{equation} \label{GenSymCritInfsm}
v \cdot \dot{\partial}_q H(\dot{q},P)
- \big( \dot{\partial}_q \wedge (\dot{v} \cdot P) \big) \cdot \partial_P H(q,P)
= ( \dot{\partial}_q \wedge \dot{W} ) \cdot \partial_P H(q,P) .
\end{equation}
\end{GenSymHam}

Conservation laws for physical motions are obtained by substituting the canonical equations of motion (\ref{CanEOM}) into Eq.~(\ref{GenSymCritInfsm}). Carrying out the same sequence of steps as in Sec.~\ref{sec:ConsLawsFromSym}, we arrive at the following Hamiltonian version of the
\begin{GenNoether}
If $f(q)=q+\eps v(q)$ is a generalized infinitesimal symmetry of $H$, i.e., if Eq.~(\ref{GenSymCritInfsm}) holds for some $D-1$-vector-valued function $W(q)$, then the solutions of the canonical equations of motion (\ref{CanEOM}) satisfy the generalized conservation law 
\begin{align} \label{GenConsLaw}
d\Gamma \cdot \partial_q \, (P \cdot v)
&= - d\Gamma \cdot \partial_q W  \hspace{17mm}{\rm for}~ D=1 \nonumber\\
(d\Gamma \cdot \partial_q) \cdot (P \cdot v)
&= (-1)^{D} (d\Gamma \cdot \partial_q) \cdot W \hspace{5mm}{\rm for}~ D>1 .
\end{align}
\end{GenNoether}

The integral form of these conservation laws is obtained readily by integrating Eq.~(\ref{GenConsLaw}) over an arbitrary connected $D$-dimensional subset $\bar{\gamma}_{\rm cl}$ of a physical motion $\gamma_{\rm cl}$, and by using the \emph{fundamental theorem of geometric calculus}, which relates an integral over $\bar{\gamma}_{\rm cl}$ to an integral over its boundary $\partial\bar{\gamma}_{\rm cl}$ \cite[Ch.~7-3]{Hestenes}. For $D=1$, we hence obtain
\begin{equation} \label{ConsLawIntD1}
P(q_2) \cdot v(q_2) - P(q_1) \cdot v(q_1) 
= - W(q_2) + W(q_1) ,
\end{equation}
where $q_1$, $q_2$ are the endpoints of a curve $\bar{\gamma}_{\rm cl}$, whereas for $D>1$, we find
\begin{equation} \label{ConsLawIntDD}
\int_{\partial \bar{\gamma}_{\rm cl}} d\Sigma \cdot (P \cdot v)
= (-1)^D \int_{\partial \bar{\gamma}_{\rm cl}} d\Sigma \cdot W ,
\end{equation}
where $d\Sigma$ is the infinitesimal element of the boundary $\partial \bar{\gamma}_{\rm cl}$.

\section{Examples}
%%%%%%%%%%%%%%%%%%%%%%%%%%%%%%%%%%%%%%
\label{sec:Examples}

Following examples illustrate how symmetries and conserved quantities can be identified in practice. In Example~\ref{sec:NonRel}, we discuss ordinary as well as generalized symmetries, while in Examples~\ref{sec:ScF} and \ref{sec:ExStr}, we content ourselves with ordinary symmetries only. We presume certain routine knowledge of the geometric algebra formalism without explaining details of all calculations.

%%%%%%%%%%%%%%%%%%%%%%%%%%%%%%%%%%%%%%
\subsection{Non-relativistic mechanics}
%%%%%%%%%%%%%%%%%%%%%%%%%%%%%%%%%%%%%%
\label{sec:NonRel}

In order to examine non-relativistic particle mechanics, we set $D=1$, and choose a time axis in the configuration space $\mathcal{C}$, i.e., a one-dimensional linear subspace spanned by a unit vector $e_t$. The position space, or $x$-space, is the orthogonal complement of the time axis. Its pseudoscalar $I_x$ satisfies  $I_x \cdot e_t = 0$ (see Appendix \ref{sec:GAGC} for details about linear subspaces). The points in $\mathcal{C}$ are decomposed accordingly as $q=t+x$, where $t\in\mathcal{G}(e_t)$ and $x\in\mathcal{G}(I_x)$. Note that the $x$-space may host coordinates of any number of particles. 

We consider the Hamiltonian
\begin{equation}
H_{\rm NR}(q,p) = p \cdot e_t + H_0(x,p_x) ,
\end{equation}
where $p_x \equiv p \cdot I_x I_x^{-1}$ is the projection of the momentum into the $x$-space, and $H_0$ represents the conventional Hamiltonian of the non-relativistic mechanics. In this example, we denote the momentum vector by $p$ instead of capital $P$. 

The generalized infinitesimal symmetry condition, Eq.~(\ref{GenSymCritInfsm}), now reads
\begin{equation} \label{NRGenSymCrit}
v \cdot \partial_x H_0 
- (e_t \cdot \partial_q v) \cdot p
- ( \dot{\partial}_x \dot{v} \cdot p ) \cdot \partial_p H_0
= e_t \cdot \partial_q W + (\partial_x W) \cdot \partial_p H_0 ,
\end{equation}
with $W(q)$ a scalar-valued function. 

Physical motions can be represented in terms of functions $x(t)$ as $\gamma_{\rm cl}=\{ t+x(t) \,|\, t \in {\rm span}\{e_t\} \}$ (for details see Ref.~\cite{ZatlCanEOM}). If we then denote $\mathsf{p}(t) \equiv p(t+x(t))$, $\mathsf{v}(t) \equiv v(t+x(t))$, and $\mathsf{W}(t) \equiv W(t+x(t))$, the conservation law (\ref{GenConsLaw}) can be rewritten in the form
\begin{equation} \label{NRConsLaw}
dt \cdot \partial_t \left[ \mathsf{p}(t) \cdot \mathsf{v}(t) + \mathsf{W}(t) \right]
= 0 ,
\end{equation}
where we used the fact that $d\Gamma \cdot \partial_q = dt \cdot \partial_t$, which follows from Eq.~(\ref{GCgdGammaPartial}).

\begin{figure} 
\includegraphics[scale=1]{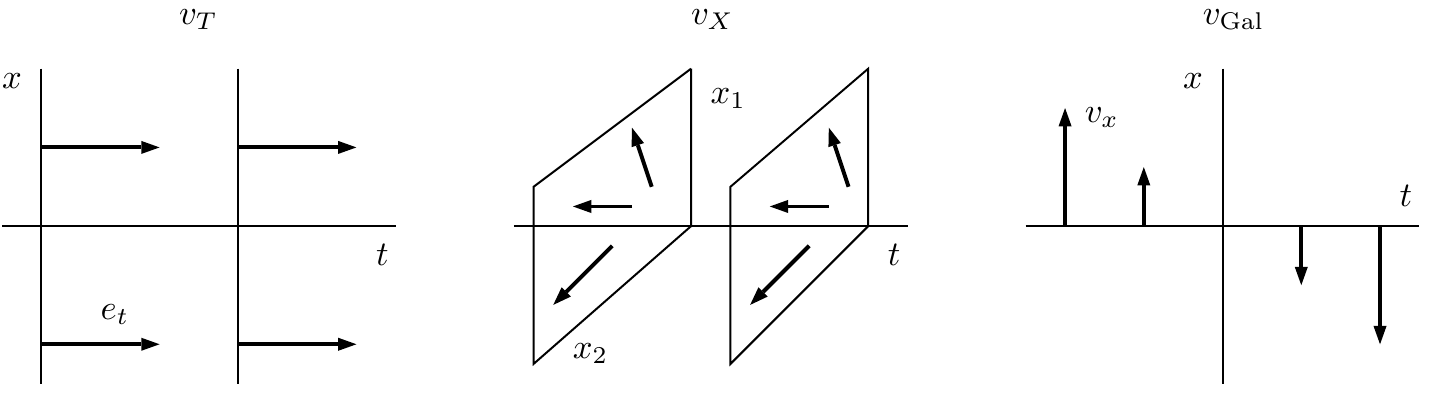}
\caption{Three examples of symmetry generators of a non-relativistic mechanical system with the relativistic Hamiltonian $H_{\rm NR}$: time translations $v_{T}$, spatial symmetries $v_{X}$ (the space is represented by the $(x_1,x_2)$-plane), and the Galilei transformations $v_{\rm Gal}$. }
\label{fig:NRGen}
\end{figure}

We will now discuss in more detail three choices of the symmetry generator $v$ (see Fig.~\ref{fig:NRGen}).

\subsubsection{Time translations}

Taking 
\begin{equation}
v_T(q) = e_t ,
\end{equation}
and $W(q)=0$, Eq.~(\ref{NRGenSymCrit}) holds true, and the ensuing conservation law (\ref{NRConsLaw}) can be cast, using the Hamiltonian constraint $H_{\rm NR} = p \cdot e_t + H_0 = 0$, as
\begin{equation}
dt \cdot \partial_t \, \mathsf{p} \cdot e_t
= - dt \cdot \partial_t \, H_0(x(t),\mathsf{p}_x(t))
= 0 .
\end{equation}
This is the conservation of energy, i.e., of the non-relativistic Hamiltonian $H_0$, corresponding to time translations $t \rightarrow t + \tau e_t$ generated by $v_T$.

\subsubsection{Spatial symmetries}

Consider
\begin{equation}
v_X(q) = v_x(x) ,
\end{equation}
where $v_x\in\mathcal{G}(I_x)$ is a generic spatial vector field, and set $W(q)=0$. The symmetry condition, Eq.~(\ref{NRGenSymCrit}), reads
\begin{equation}
v_x \cdot \partial_x H_0 - (\dot{\partial}_x \dot{v}_x \cdot p_x) \cdot \partial_p H_0 
= 0 ,
\end{equation}
where the left-hand side coincides with the standard Poisson bracket of non-relativistic Hamiltonian mechanics $\{ H_0 , v_x \cdot p_x \}$. The scalar product $v_x \cdot p_x$ is the corresponding conserved quantity, which satisfies
\begin{equation}
dt \cdot \partial_t \big(\mathsf{p}_x(t) \cdot \mathsf{v}_x(t) \big) = 0 .
\end{equation}

For example, if $v_x$ is a constant vector field such that $v_x \cdot \partial_x H_0 = 0$, then $v_X$ is indeed a symmetry generator, and the corresponding conserved quantity is simply the $v_x$-component of the momentum.

\subsubsection{Galilei transformations}

The Galilei transformations are included to illustrate the concept of generalized symmetries and the generalized Noether theorem of Sec.~\ref{sec:GenSym}. We assume the non-relativistic Hamiltonian in the form
\begin{equation} \label{NRHamKinPot}
H_0 = p_x^2 + V(x) ,
\end{equation}
and consider a transformation $q \mapsto q'$,
\begin{equation}
q' = q - q \cdot e_t \, v_x ,
\end{equation}
or, more explicitly, 
\begin{equation}
x' = x - v_x \, e_t \cdot t 
\quad,\quad
t' = t ,
\end{equation}
where $v_x$ is a constant $x$-space vector.
Plugging the associated infinitesimal generator
\begin{equation}
v_{\rm Gal}(q) = -q \cdot e_t \, v_x 
\end{equation}
into Eq.~(\ref{NRGenSymCrit}), and taking into account the form of the Hamiltonian (\ref{NRHamKinPot}), we obtain the symmetry condition
\begin{equation} \label{NRSymCondGal}
- q \cdot e_t \, v_x \cdot \partial_x V 
+ p_x \cdot v_x
= e_t \cdot \partial_q W + p_x \cdot \partial_x W .
\end{equation}

Let us assume that the potential is such that
\begin{equation} \label{NRPotAss}
v_x \cdot \partial_x V = 0 .
\end{equation}
Then, Eq.~(\ref{NRSymCondGal}) holds if we set, for example,
\begin{equation}
W(q) = v_x \cdot x ,
\end{equation}
and the conservation law (\ref{NRConsLaw}) renders the form
\begin{equation}
dt \cdot \partial_t \left[ x(t) - t \cdot e_t \, \mathsf{p}_x(t) \right] \cdot v_x = 0 .
\end{equation}

%When $x$ gathers coordinates of $n$ particles, $x = x^{(1)}+\ldots+x^{(n)}$, and the potential 
%\begin{equation}
%V(x)=\sum_{j<k}V_2(x_j-x_k)
%\end{equation}
%describes a pairwise interparticle interaction, 

%%%%%%%%%%%%%%%%%%%%%%%%%%%%%%%%%%%%%%
\subsection{Scalar field theory}
%%%%%%%%%%%%%%%%%%%%%%%%%%%%%%%%%%%%%%
\label{sec:ScF}

In this example, we split the configuration space $\mathcal{C}$ into a $D$-dimensional spacetime with a unit pseudoscalar $I_x$ (we will assume $D>1$), and its $N$-dimensional orthogonal complement, the internal space, or the space of fields, with an orthonormal basis $\{ e_a \}_{a=1}^N$ and a unit pseudoscalar $I_y$. The points in $\mathcal{C}$ then have a natural decomposition $q=x+y$.

We assume the following form of the Hamiltonian:
\begin{equation} \label{HamDWform}
H(q,P) = P \cdot I_x + H_{\rm DW}(q,P) ,
\end{equation}
where $H_{\rm DW}$ is the so-called De Donder-Weyl Hamiltonian \cite{DeDonder,Weyl,Kanat1998}, which satisfies 
\begin{equation} \label{DWcond}
I_x \cdot \partial_P H_{\rm DW} = 0 \quad,\quad
(e_b \wedge e_a) \cdot \partial_P H_{\rm DW} = 0 
\quad (\forall a,b=1,\ldots,N) .
\end{equation}
Geometrically, these conditions mean that $H_{\rm DW}$ depends only on the components of the momentum $D$-vector that are composed of one vector from the $y$-space, and $D-1$ vectors from the $x$-space. Note that in Ref.~\cite{ZatlCanEOM} we treated in detail the $N=1$-case of this field theory.

In order to make contact with the standard theory of fields as functions defined on the spacetime, we represent the motions as $\gamma=\{ x+y(x)\,|\,x\in\Omega \}$, where $\Omega$ is a spacetime domain. The surface element of $\gamma$ is related to the oriented spacetime element $dX=|dX|I_x$ via Formula (\ref{GCgdGamma}),
\begin{equation} \label{SFdGamma}
d\Gamma = dX + (dX \cdot \partial_x) \wedge y ,
\end{equation}
where the terms with more than one $y$ vanish as a consequence of the second of the conditions (\ref{DWcond}), and the first canonical equation (\ref{CanEOM1}), which for the Hamiltonian (\ref{HamDWform}) reads
\begin{equation} \label{CanEOM1DW}
d\Gamma = \lambda I_x + \lambda \, \partial_P H_{\rm DW} .
\end{equation}
We may also assume that the classical momentum satisfies
\begin{equation} \label{SFMomAss}
P \cdot (e_a \wedge e_b) = 0 \quad (\forall a,b) ,
\end{equation}
as this condition has no effect on the classical motions.

Equation (\ref{GCgdGammaPartial}) can be used to ``pull" the conservation law (\ref{ConsLaw}) down onto the spacetime to find the standard form of the continuity equation, and relate the conserved multivectors $P \cdot v$ to the vectorial Noether currents. For this purpose we  define $\mathsf{P}(x) \equiv P(x+y(x))$, $\mathsf{v}(x) \equiv v(x+y(x))$, and calculate
\begin{align}
(d\Gamma \cdot \partial_q) \cdot (P \cdot v)
&= \left[ dX \cdot \partial_x + \big( (dX \cdot \partial_x) \cdot \dot{\partial}_x \big) \wedge \dot{y} \right] \cdot (\mathsf{P} \cdot \mathsf{v}) \nonumber\\
&= (-1)^{D-1} (\partial_x \cdot dX) \cdot \left[ \mathsf{P} \cdot \mathsf{v} 
+ \dot{\partial}_x \wedge \big( \dot{y} \cdot (\mathsf{P} \cdot \mathsf{v}) \big) \right] \nonumber\\
&= |dX| (-1)^D \partial_x \cdot j(x) ,
\end{align}
where we have denoted
\begin{equation} \label{NoetherCur}
j(x) \equiv - I_x \cdot \left[ \mathsf{P} \cdot \mathsf{v} + \dot{\partial}_x \wedge \big( \dot{y} \cdot (\mathsf{P} \cdot \mathsf{v}) \big) \right] .
\end{equation}
This is the traditional \emph{Noether current} corresponding to a symmetry generated by a vector field $v$. In view of Eq.~(\ref{ConsLaw}), it satisfies the continuity equation
\begin{equation}
\partial_x \cdot j(x) = 0 .
\end{equation}

To discuss concrete examples of symmetries and corresponding conserved quantities, we will specialize our Hamiltonian (\ref{HamDWform}) to an $x$-independent scalar-field Hamiltonian (the name is given justice in Eqs.~(\ref{ActionSF}) and (\ref{LagrSF}) below)
\begin{equation} \label{HamScField}
H_{\rm SF}(q,P) = P \cdot I_x + \frac{1}{2} \sum_{a=1}^N \big( I_x \cdot (P \cdot e_a) \big)^2 + V(y) .
\end{equation}
Eq.~(\ref{CanEOM1DW}) now reads
\begin{equation} \label{CanEOM1SF}
d\Gamma
= \lambda I_x + \lambda \sum_{a=1}^N e_a \wedge (e_a \cdot \rev{P}) ,
\end{equation}
where the assumption (\ref{SFMomAss}) has been taken into account. Comparing this result with Eq.~(\ref{SFdGamma}), we find immediately that
\begin{equation} \label{SFlambda}
\lambda = |dX| ,
\end{equation}
and
\begin{equation}
(I_x \cdot \partial_x) \wedge y = \sum_{a=1}^N e_a \wedge (e_a \cdot \rev{\mathsf{P}}) ,
\end{equation}
where $\rev{\mathsf{P}}$ denotes the reversion of $\mathsf{P}$.
The latter equation is equivalent with
\begin{equation} \label{SFMomSubs}
\rev{I}_x \cdot \partial_x \phi_a = \mathsf{P} \cdot e_a ,
\end{equation}
where $\phi_a \equiv e_a \cdot y$ denotes the $a$-th component of the field $y(x)$.

At this point it is worth to remark that the extended action (\ref{ActionAugm}) for the Hamiltonian $H_{\rm SF}$ can be cast, using Eqs.~(\ref{SFdGamma}) and (\ref{SFlambda}), as an integral over the spacetime domain $\Omega$,
\begin{align}
\mathcal{A}_{\rm SF}
&= \int_\Omega \left\{ \mathsf{P} \cdot \left[ dX + (dX \cdot \partial_x) \wedge y \right] - |dX| H_{\rm SF} \right\} \nonumber\\
&= \int_\Omega |dX| \left\{ (I_x \cdot \dot{\partial}_x) \cdot (\dot{y} \cdot \mathsf{P}) 
- \frac{1}{2} \sum_{a=1}^N \big( I_x \cdot (\mathsf{P} \cdot e_a) \big)^2 - V(y) \right\} .
\end{align}
Eliminating the momentum by virtue of Eq.~(\ref{SFMomSubs}) and employing the identity (\ref{GAprojInner}) with $A=I_x$ (keeping in mind that in Euclidean spaces $I_x^{-1}=\rev{I}_x$) then yields
\begin{equation} \label{ActionSF}
\mathcal{A}_{\rm SF}
= \int_\Omega \mathcal{L}_{\rm SF}(\phi_a,\partial_x \phi_a) \, |dX| ,
\end{equation}
where
\begin{equation} \label{LagrSF}
\mathcal{L}_{\rm SF}(\phi_a,\partial_x \phi_a)
= \frac{1}{2} \sum_{a=1}^N (\partial_x \phi_a)^2 - V(y)
\end{equation}
is the usual Lagrangian of an $N$-component scalar field $y=(\phi_1,\ldots,\phi_N)$.

\begin{figure} 
\includegraphics[scale=1]{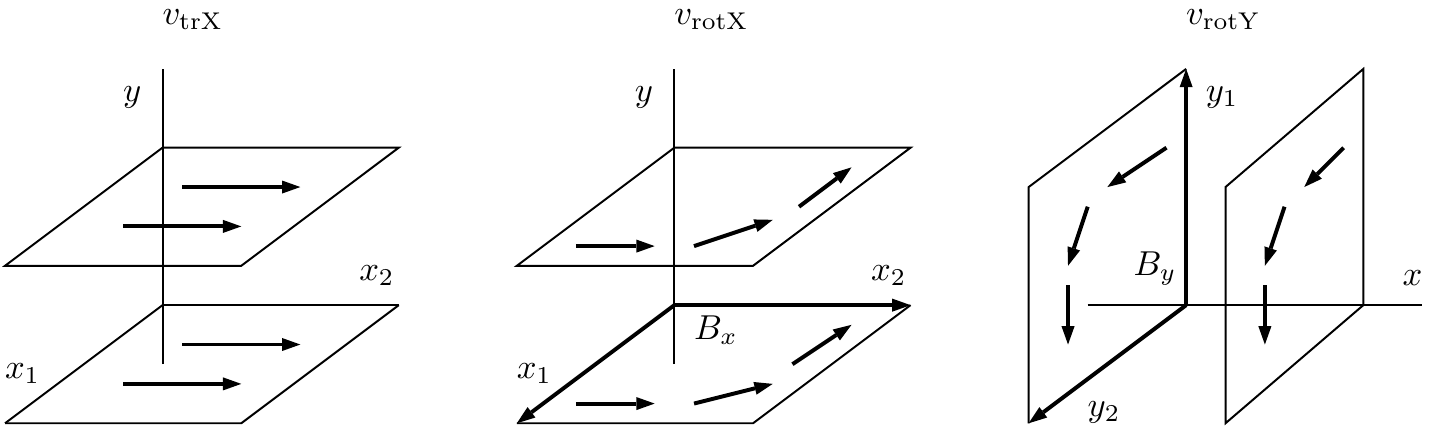}
\caption{The generators of symmetries of the scalar field Hamiltonian $H_{\rm SF}$: spacetime translations $v_{\rm trX}$, spacetime rotations $v_{\rm rotX}$, and field-space rotations $v_{\rm rotY}$. The spacetime or the field space are conveniently depicted as two-dimensional planes $(x_1,x_2)$ or $(y_1,y_2)$, respectively.}
\label{fig:SFGen}
\end{figure}

We will now show that the scalar-field Hamiltonian $H_{\rm SF}$ enjoys some well known symmetries (depicted in Fig.~\ref{fig:SFGen}), and exploit the corresponding conserved currents.

\subsubsection{Translations in spacetime}

For global spacetime translations
\begin{equation}
f_{\rm trX}(q) = q + v_x ,
\end{equation}
where $v_x$ is a constant spacetime vector, the differential mapping is trivial, 
\begin{equation}
\underline{f}(a) = a ,
\end{equation}
and so is the adjoint,
\begin{equation}
\overline{f^{-1}}(P) = P .
\end{equation}
The vector $v_x$ is at the same time the generator of translations,
\begin{equation}
v_{\rm trX}(q) = v_x ,
\end{equation} 
as can be ascertained by calculating $e^{v_x \cdot \partial_q} q = f_{\rm trX}(q)$ (recall the Lie series, Eq.~(\ref{LieSeries})).

The transformation $f_{\rm trX}$ is according to Eq.~(\ref{SymCrit}) a symmetry of the Hamiltonian $H_{\rm SF}$, since $H_{\rm SF}$ does not depend on $x$. The conserved quantity $P \cdot v_{\rm trX}$ is related to the Noether current $j_{\rm trX}(x)$ via Eq.~(\ref{NoetherCur}). Explicitly,
\begin{align}
j_{\rm trX} 
&= - I_x \cdot \left[ \mathsf{P} \cdot v_x + \big( (\mathsf{P} \cdot \dot{y}) \wedge  \dot{\partial}_x \big) \cdot v_x - v_x \cdot \dot{\partial}_x \, \mathsf{P} \cdot \dot{y} \right] 
\nonumber\\
&= - v_x \left[ \mathsf{P} \cdot I_x + (\mathsf{P} \cdot \dot{y}) \cdot  (\dot{\partial}_x \cdot I_x) \right]
+ v_x \cdot \dot{\partial}_x \, I_x \cdot (\mathsf{P} \cdot \dot{y}) ,
\end{align}
where we have used the identities (\ref{GAident0}) and (\ref{GAident1}), and the fact that $I_x \wedge v_x = 0$. Substituting now for $P \cdot I_x$ from the Hamiltonian constraint $H_{\rm SF} = 0$, and for $P \cdot e_a$ from Eq.~(\ref{SFMomSubs}), and using Eq.~(\ref{GAprojInner}), we arrive at
\begin{align} \label{SFEnMomTensor}
j_{\rm trX}(x;v_x)
&= - v_x \left[ \frac{1}{2} \sum_{a=1}^N (\partial_x \phi_a)^2 - V(y) \right] + \sum_{a=1}^N ( v_x \cdot \partial_x \phi_a ) (\partial_x \phi_a)
\nonumber\\
&= - v_x \mathcal{L}_{\rm SF} + \sum_{a=1}^N ( v_x \cdot \partial_x \phi_a ) \frac{\partial \mathcal{L}_{\rm SF}}{\partial (\partial_x \phi_a)} .
\end{align}
This is the energy-momentum tensor of a scalar field with Lagrangian (\ref{LagrSF}).
In its natural geometric interpretation, $j_{\rm trX}$ is an $x$-dependent linear mapping of spacetime vectors $v_x$ to spacetime vectors $j_{\rm trX}(x;v_x)$.

\subsubsection{Rotations in spacetime}

A spacetime rotation about a point $x_0$ is defined
\begin{equation} \label{RotXdef}
f_{\rm rotX}(q) = x_0 + R_x (q - x_0) \rev{R}_x 
~~~,~~~
R_x = e^{-B_x/2} ,
\end{equation}
where $B_x$ is a constant spacetime bivector, $B_x \in \mathcal{G}(I_x)$, and $R_x$ is the corresponding rotor. The associated differential mapping is readily obtained,
\begin{equation}
\underline{f_{\rm rotX}}(a) 
= a \cdot \partial_q f_{\rm rotX}(q)
= R_x a \rev{R}_x ,
\end{equation}
and the transformation rule for the momentum is found,
\begin{equation}
\overline{f_{\rm rotX}^{-1}}(P) = R_x P \rev{R}_x .
\end{equation}
(The geometric algebra implementation of rotations is discussed in detail in Appendix~\ref{sec:GARot}.)

By expanding the right-hand side of definition (\ref{RotXdef}) according to Eq.~(\ref{GARotExpansion}), and comparing with the Lie series, Eq.~(\ref{LieSeries}), we find the infinitesimal generator of $f_{\rm rotX}$,
\begin{equation} \label{SFGenRotX}
v_{\rm rotX}(q) = (q - x_0) \cdot B_x = (x - x_0) \cdot B_x .
\end{equation}

In order to show that $f_{\rm rotX}$ is a symmetry of $H_{\rm SF}$, we realize that $R_x I_x \rev{R}_x = I_x$ and $R_x e_a \rev{R}_x = e_a$ (see Eq.~(\ref{GARotSubspace}) and the discussion around), and calculate
\begin{align}
H_{\rm SF}(f_{\rm rotX}(q), \overline{f_{\rm rotX}^{-1}}(P))
&= (R_x P \rev{R}_x) \cdot I_x + \frac{1}{2} \sum_{a=1}^N \big[ I_x \cdot \big((R_x P \rev{R}_x) \cdot e_a \big) \big]^2 + V(y) \nonumber\\
%%%%%%
%&= (R_x P \rev{R}_x) \cdot (R_x I_x \rev{R}_x) + \frac{1}{2} \sum_{a=1}^N \big[ (R_x I_x \rev{R}_x) \cdot \big( (R_x P \rev{R}_x) \cdot (R_x e_a \rev{R}_x) \big) \big]^2 + V(y) \nonumber\\
%%%%%%
&= P \cdot I_x + \frac{1}{2} \sum_{a=1}^N \big( I_x \cdot (P \cdot e_a) \big)^2 + V(y) = H_{\rm SF}(q,P) .
\end{align}
Since $v_{\rm rotX}$ is a spacetime vector, it is easy to find an explicit relation between $P \cdot v_{\rm rotX}$ and the corresponding Noether current $j_{\rm rotX}$. We simply replace in Eq.~(\ref{SFEnMomTensor}) $v_x$ by $\mathsf{v}_{\rm rotX}$:
\begin{equation} \label{SFAngMomTensor}
j_{\rm rotX}(x;B_x,x_0) 
= j_{\rm tr}\big(x; \mathsf{v}_{\rm rotX} \big)
= j_{\rm tr}\big(x;(x-x_0)\cdot B_x \big) .
\end{equation}
This is the angular momentum tensor corresponding to the energy-momentum tensor $j_{\rm tr}$. Geometrically, $j_{\rm rotX}$ is an $x$-dependent linear mapping, with a parameter $x_0$, that maps spacetime bivectors $B_x$ to spacetime vectors $j_{\rm rotX}(x;B_x,x_0)$ (c.f. Ch.~13.1 in Ref.~\cite{DoranLas}).

\subsubsection{Rotations in field space}

Finally, let us consider rotations of the form
\begin{equation}
f_{\rm rotY}(q) = R_y q \rev{R}_y 
~~~,~~~
R_y = e^{-B_y/2} ,
\end{equation}
where $B_y$ is a constant bivector from the field space, $B_y \in \mathcal{G}(I_y)$. The differential reads
\begin{equation}
\underline{f_{\rm rotY}}(a) 
= a \cdot \partial_q f_{\rm rotY}(q)
= R_y a \rev{R}_y ,
\end{equation}
and the momentum transforms as
\begin{equation}
\overline{f_{\rm rotY}^{-1}}(P) = R_y P \rev{R}_y .
\end{equation}
The generator of field-space rotations is found in the same way as the generator of spacetime rotations (\ref{SFGenRotX}), 
\begin{equation}
v_{\rm rotY}(q) = q \cdot B_y = y \cdot B_y .
\end{equation}

The Hamiltonian $H_{\rm SF}$ transforms under $f_{\rm rotY}$ as follows (note that $R_y I_x \rev{R}_y = I_x$):
\begin{equation} \label{SFHamRotY}
H_{\rm SF}(f_{\rm rotY}(q), \overline{f_{\rm rotY}^{-1}}(P))
= P \cdot I_x + \frac{1}{2} \sum_{a=1}^N \big[ I_x \cdot \big(P \cdot (\rev{R}_y e_a R_y) \big) \big]^2 + V(R_y y \rev{R}_y) .
\end{equation}
If we assume that $V(R_y y \rev{R}_y) = V(y)$, which is fulfilled, for example, when the potential $V$ depends only on $y^2 = \sum_a \phi_a^2$, then the right-hand side of Eq.~(\ref{SFHamRotY}) is equal to $H_{\rm SF}(q,P)$, and hence $f_{\rm rotY}$ is a symmetry of $H_{\rm SF}$.  Note that the second term in $H_{\rm SF}$ is invariant under a change of the orthonormal basis of the $y$-space, $e_a \rightarrow e_a'=\rev{R}_y e_a R_y$, as can be easily ascertained.

The vector field $v_{\rm rotY}$ lies entirely in the $y$-space. Therefore, owing to the assumption (\ref{SFMomAss}), the second term in expression (\ref{NoetherCur}) for the Noether current drops out, and we obtain
\begin{equation}
j_{\rm rotY}
= -I_x \cdot (\mathsf{P} \cdot \mathsf{v}_{\rm rotY})
= - \sum_{a=1}^N I_x \cdot (\mathsf{P} \cdot e_a) \, (e_a \wedge y) \cdot B_y .
\end{equation}
A substitution for the momentum from Eq.~(\ref{SFMomSubs}) then yields
\begin{equation}
j_{\rm rotY}(x;B_y)
= \dot{\partial}_x \, (y \wedge \dot{y}) \cdot B_y
= \sum_{a,b=1}^N (e_a \wedge e_b) \cdot B_y \, \phi_a \partial_x \phi_b .
\end{equation}
The Noether current $j_{\rm rotY}$ is an $x$-dependent linear mapping of field-space bivectors $B_y$ to spacetime vectors $j_{\rm rotY}(x;B_y)$.

%%%%%%%%%%%%%%%%%%%%%%%%%%%%%%%%%%%%%%
\subsection{String theory}
%%%%%%%%%%%%%%%%%%%%%%%%%%%%%%%%%%%%%%
\label{sec:ExStr}

Let us consider the Hamiltonian
\begin{equation}
H_{\rm Str} = \frac{1}{2}(|P|^2 - \Lambda^2) ,
\end{equation}
where $\Lambda >0$ is a scalar constant, and $|P|$ is the magnitude of $P$ (see the definition (\ref{GAMagnitude})). In Ref.~\cite{ZatlCanEOM}, it has been shown that this Hamiltonian describes the dynamics of a relativistic particle (for $D=1$), a Nambu-Goto bosonic string (for $D=2$), or a higher-dimensional membrane (for $D>2$) that propagates in a Euclidean spacetime $\mathcal{C}$. The corresponding worldlines (or worldsheets) are identified with the motions $\gamma$.

The infinitesimal symmetry condition, Eq.~(\ref{SymCritInfsm}), now reads
\begin{equation} \label{StrSym}
\big( \dot{\partial}_q \wedge (\dot{v} \cdot P) \big) \cdot \rev{P}
= 0 ,
\end{equation}
which is to be satisfied for all constant $D$-vectors $P$. Observe that the left-hand side is equal to
\begin{equation}
\frac{1}{2} \left[ \dot{\partial}_q \wedge (\dot{v} \cdot P) + \dot{v} \wedge (\dot{\partial}_q \cdot P) \right] \cdot \rev{P}
= \frac{1}{2} \sum_{j=1}^{N+D} ( \partial_q v \cdot e_j + e_j \cdot \partial_q v ) \cdot \big( (e_j \cdot P) \cdot \rev{P} \big) ,
\end{equation}
where the $e_j$'s form an arbitrary orthonormal basis.
%Differentiation with respect to $P$ yields
%\begin{equation}
%\dot{\partial}_q \wedge (\dot{v} \cdot P)
%+ (P \cdot \partial_q) \wedge v = 0 ,
%\end{equation}
Eq.~(\ref{StrSym}) is therefore solved by a vector field $v$, for which 
\begin{equation} \label{SymGenStr}
a \cdot \partial_q v = - \partial_q v \cdot a 
\end{equation}
holds for all constant vectors $a$. 
%In view of definitions (\ref{GAdifMap}) and (\ref{GCadjoint}), we can write succinctly $\uline{v}=-\oline{v}$, i.e., $\uline{v}$ is a skew-symmetric linear mapping.
Taking the curl, the right-hand side vanishes, and we find 
\begin{equation}
a\cdot\partial_q \, \partial_q \wedge v = 0 ,
\end{equation}
which implies
\begin{equation}
\partial_q \wedge v = 2 B_0 ,
\end{equation}
where $B_0$ is a constant bivector. Moreover, note that from Eq.~(\ref{SymGenStr}) it follows that 
\begin{equation}
a \cdot ( \partial_q \wedge v ) = 2 a \cdot \partial_q v ,
\end{equation}
and hence
\begin{equation}
a \cdot \partial_q v - a \cdot B_0 = a \cdot \partial_q (v - q \cdot B_0) = 0 .
\end{equation}
From the last equality we finally obtain the expression for the symmetry generator $v$:
\begin{equation} \label{TrRotGen}
v(q) = q \cdot B_0 + v_0 ,
\end{equation}
where $v_0$ is a constant vector.

The vector field $v$ is composed of two terms --- the translation generator 
\begin{equation}
v_{\rm tr}(q) = v_0 ,
\end{equation}
and the rotation generator
\begin{equation}
v_{\rm rot}(q) = q \cdot B_0 .
\end{equation}
Corresponding finite symmetry transformations can be obtained directly from the Lie series, Eq.~(\ref{LieSeries}), and read (for $\tau = 1$)
\begin{equation}
f_{\rm tr}(q) = q + v_0 ,
\end{equation}
and
\begin{equation}
f_{\rm rot}(q) 
= q + q \cdot B_0 + \frac{1}{2!} \big( q \cdot B_0 \big) \cdot B_0 + \ldots
= e^{- B_0/2} \, q \, e^{B_0/2} .
\end{equation}
(The last equality follows from Formula~(\ref{GARotExpansion}).)

In the string theory example in Ref.~\cite{ZatlCanEOM}, the momentum has been related to the unit pseudoscalar of a surface $\gamma$ by $P=\pm \Lambda \rev{I}_\gamma$ (the sign is equal to $\lambda/|\lambda|$). The conserved quantities are then expressed as
\begin{equation}
P \cdot v = \pm \Lambda \rev{I}_\gamma \cdot v .
\end{equation}

Finally, let us say a few general words about the conserved quantities that correspond to the symmetry generator (\ref{TrRotGen}). In mechanics, i.e., for $D=1$, the translation symmetry induces conservation of the $v_0$-component of the momentum vector $P$, $P\cdot v_0$. On the other hand, the rotation symmetry ensures conservation of $P \cdot (q \cdot B_0) = (P \wedge q) \cdot B_0$, i.e., the $B_0$-component of the angular momentum bivector (see \cite[Ch~3.1]{DoranLas} for further explanation). 

In field theory, i.e., for $D>1$, the quantities $P\cdot v_0$ and $P\cdot (q\cdot B_0)$ establish a natural generalization of these mechanical notions. In particular, as we have seen in Example~\ref{sec:ScF}, $P \cdot v_0$ is related to the energy-momentum tensor of the scalar field theory, while $P\cdot (q\cdot B_0)$ relates to its angular-momentum tensor and to the internal currents.

%%%%%%%%%%%%%%%%%%%%%%%%%%%%%%%%%%%%%%
\section{Conclusion and outlook}
%%%%%%%%%%%%%%%%%%%%%%%%%%%%%%%%%%%%%%

We have discussed transformations of the configuration space of partial observables in the Hamiltonian constraint approach to classical field theories. The symmetry transformations were identified as the mappings satisfying Eq.~(\ref{SymCrit}), or its infinitesimal form, Eq.~(\ref{SymCritInfsm}). Together with the canonical equations of motion, symmetries were shown to imply conservation laws, Eq.~(\ref{ConsLaw}), with $D-1$-vector-valued conserved quantities $P \cdot v$, where $v$ was a symmetry-generating vector field. We have therefore established a Hamiltonian field-theoretic version of the Noether theorem. A slight generalization of these results was then derived in Sec.~\ref{sec:GenSym}.

Three examples were included to illustrate the general theory and put our results in context with a well-established treatment of symmetries. In the case of scalar field theory, in particular, we have shown how our conservation laws relate to the traditional continuity equations, and how the energy-momentum tensor as well as internal Noether currents are expressed in terms of the conserved multivector $P \cdot v$.

Our treatment of symmetries and conservation laws should be compared with standard Hamiltonian treatments of non-relativistic mechanical systems (see, e.g., Ch.~9 in monograph \cite{Goldstein}). There, an emphasis is put on the theory of canonical transformations and Poisson brackets, and the Noether theorem is usually mentioned as its corollary. On the contrary, we decided to get directly to the point and establish the Noether theorem without actually introducing the canonical transformation machinery. Our approach is relatively fast in providing physically relevant results. Still, a field-theoretic treatment of canonical transformations and Poisson brackets could provide a valuable insight into the structure of Hamiltonian field theories, and it clearly deserves a separate article to address these issues properly. In this respect, we would like to mention Refs.~\cite{Struckmeier} and \cite{Kanat1997}, which deal with the De Donder-Weyl Hamiltonian field theory, and which could provide some useful inspiration.

%%%%%%%%%%%%%%%%%%%%%%%%%%%%%%%%%%%%%%
\subsection*{Acknowledgement}
%%%%%%%%%%%%%%%%%%%%%%%%%%%%%%%%%%%%%%

The author received support from
Czech Science Foundation (GA\v{C}R), Grant GA14-07983S, and Deutsche Forschungsgemeinschaft (DFG), Grant KL 256/54-1.

\appendix
%%%%%%%%%%%%%%%%%%%%%%%%%%%%%%%%%%%%%%
\section{Elements of geometric algebra and calculus} \label{sec:GAGC}
%%%%%%%%%%%%%%%%%%%%%%%%%%%%%%%%%%%%%%

%A concise introduction into geometric algebra and calculus has been given in our previous article \cite{ZatlCanEOM}. A lot more can be found in the monographs \cite{Hestenes, DoranLas}. 

The purpose of this appendix is to fix the basic geometric algebra notation, and recall some useful formulas, which find applications in the main text. We shall devote special attention to the way geometric algebra handles rotations, and to the theory of diffeomorphism and induced mappings.

Let $V$ be a finite-dimensional real vector space, and consider its geometric (or Clifford) algebra $\mathcal{G}(V)$, with the geometric product denoted by an empty symbol, i.e., a juxtaposition. The inner and outer products, denoted by ``$\,\cdot\,$" and ``$\wedge$", are defined to be, respectively, the lowest and highest-grade elements in the geometric product of two multivectors, with the exception that the inner product with a scalar is set to zero. We will assume that the inner product of vectors is positive-definite.

If $a$ is a vector, and $A_r$ and $B_s$ multivectors of grade $r$ and $s$, respectively, then the following identities hold (proofs can be found in \cite[Ch.~1-1]{Hestenes}):
\begin{align} \label{GAident0}
A_r \cdot B_s &= (-1)^{r (s-1)} B_s \cdot A_r ~~~{\rm for}~r \leq s, \nonumber\\
(A_r \cdot B_s) \cdot a &= A_r \cdot (B_s \cdot a) \,~~~~~~~~~~{\rm for}~r < s, \nonumber\\
B_s \cdot (a \wedge A_r) = (B_s \cdot a) \cdot A_r &= (-1)^r (B_s \cdot A_r) \cdot a ~~~{\rm for}~ s > r \geq 1.
\end{align}
In addition,
\begin{align} \label{GAident1}
a \cdot (A_r \wedge B_s) &= (a \cdot A_r) \wedge B_s + (-1)^r A_r \wedge (a \cdot B_s) ,
\nonumber \\  
a \wedge (A_r \cdot B_s) &= (a \cdot A_r) \cdot B_s + (-1)^r A_r \cdot (a \wedge B_s) ~~~~~ {\rm for}~ s \geq r > 1 .
\end{align}

A multivector $A$ is called a \emph{blade}, if it admits a decomposition into an outer product of vectors, $A = \blade{a}{D}$. (These vectors can always be chosen orthogonal: $a_j \cdot a_k = \delta_{jk}$.) To any blade $A$ corresponds a $D$-dimensional vector subspace, spanned by the vectors $a_j$, and a geometric algebra $\mathcal{G}(A)$ built upon it. The blade $A$ is called the \emph{pseudoscalar} of this vector subspace. A vector $a$ is an element of $\mathcal{G}(A)$, i.e., it can be written as a linear combination of the $a_j$'s, if and only if $a \wedge A = 0$.

The magnitude of a general $D$-vector $A$ is defined
\begin{equation} \label{GAMagnitude}
|A| := \sqrt{ \rev{A} \cdot A } ,
\end{equation}
where $\rev{A}$ reverses the order of vectors in $A$. For example, if $A = a_1 \ldots a_D$ is a blade decomposed into orthogonal vectors, then $|A|=\sqrt{\rev{A}A}=\sqrt{a_1^2 \ldots a_D^2}$. As a consequence, any blade $A$ has an inverse $A^{-1}=\rev{A}/|A|^2$. (Here we rely on the fact that there are no null vectors in $V$.) By counting the number of exchanges of the factors in $A$ we find explicitly $\rev{A} = (-1)^{D(D-1)/2} A$.

Any vector $v \in V$ can be decomposed into two parts,
\begin{equation}
v = v A A^{-1} = v \cdot A A^{-1} + v \wedge A A^{-1}
= (v \cdot A) \cdot A^{-1} + (v \wedge A) \cdot A^{-1} ,
\end{equation}
where the first is the orthogonal \emph{projection} of $v$ onto a blade $A$, while the second is the \emph{rejection} of $v$ from $A$. Indeed, the projection lies entirely in $\mathcal{G}(A)$, while the rejection is orthogonal to every vector in $\mathcal{G}(A)$ (see \cite[Ch.~1-2]{Hestenes}). In general, 
\begin{equation}
B \cdot A A^{-1} = (B \cdot A) \cdot A^{-1} = A^{-1} \cdot (A \cdot B)
\end{equation}
is the projection of an arbitrary multivector $B$, whose grades are greater than $0$ and lesser than $D$ (the grade of $A$), onto $\mathcal{G}(A)$. Apparently, $B$ is an element of $\mathcal{G}(A)$ if and only if $B \cdot A = B A$.

It is useful to realize that if $a$ and $b$ are vectors, and $a \in \mathcal{G}(A)$, then
\begin{equation} \label{GAprojInner}
(A \cdot a) \cdot (b \cdot A^{-1})
= (a \cdot A) \cdot (A^{-1} \cdot b)
= \big( (a \cdot A) \cdot A^{-1} \big) \cdot b
= a \cdot b .
\end{equation}

%%%%%%%%%%%%%%%%%%%%%%%%%%%%%%%%%%%%%%
\subsection{Rotations} \label{sec:GARot}
%%%%%%%%%%%%%%%%%%%%%%%%%%%%%%%%%%%%%%

Rotations can be handled with the geometric algebra in a particularly elegant and efficient way. By definition, they are linear transformations $a \mapsto a' \equiv \uline{h}(a)$ on the vector space $V$ that preserve the inner product of vectors: 
\begin{equation}
a \cdot b = \uline{h}(a) \cdot \uline{h}(b) = a \cdot \oline{h}\uline{h}(b) ,
\end{equation}
where $\oline{h}$ denotes the adjoint transformation, i.e., in matrix representation, the transposed matrix. (We usually omit brackets when composing linear functions.) Clearly, the inverse rotation coincides with the adjoint: $\uline{h}^{-1}=\oline{h}$. 

Any rotation can be represented in the form
\begin{equation} \label{GARotDef}
a' = \uline{h}(a) = R a \rev{R} ,
\end{equation}
where
\begin{equation} \label{GARotorDef}
R = e^{-B/2} = \sum_{k=0}^\infty \frac{(-B/2)^k}{k!} 
\end{equation}
is a \emph{rotor}, and $B$ a bivector that fully characterizes the rotation \cite[Ch.~3-5]{Hestenes}. (Powers in the series are taken with the geometric product.) Rotors are multivectors composed of even-grade terms. Note that from $\rev{B}=-B$ it follows that 
\begin{equation}
\rev{R}=e^{B/2} ,
\end{equation}
and hence the characteristic property of rotors
\begin{equation} \label{GARotorNorm}
\rev{R} R = R \rev{R} = 1 .
\end{equation} 
This can be used to quickly find the inverse rotation, 
\begin{equation}
\uline{h}^{-1}(a')=\rev{R} a' R .
\end{equation}

Let us now discuss some properties of rotations that can be deduced from the above rotor representation.
First, notice that the series expansion of Eq.~(\ref{GARotDef}),
\begin{equation} \label{GARotExpansion}
a' = a + a \cdot B + \frac{1}{2!} (a \cdot B) \cdot B + \ldots ,
\end{equation}
confirms that $a'$ is indeed a vector, since inner product between a vector and a bivector always results in a vector (see also \cite[Ch.~4.2]{DoranLas}). 

A generic product of vectors transforms under rotations as
\begin{equation}
a'_1 \ldots a'_r = R a_1 \rev{R} \ldots R a_r \rev{R} = R a_1 \ldots a_r \rev{R} ,
\end{equation}
where we have made use of the property (\ref{GARotorNorm}). Rotations can be therefore naturally extended to linear grade-preserving transformations on the entire geometric algebra $\mathcal{G}(V)$:
\begin{equation}
A' = \uline{h}(A) = R A \rev{R} ,
\end{equation}
where $A$ is a generic multivector. 

That transformation (\ref{GARotDef}) preserves the inner product of vectors is now easy to check. Recalling the definition $a \cdot b = (ab+ba)/2$, we have
\begin{equation}
a' \cdot b' = R \, a \cdot b \, \rev{R} = a \cdot b .
\end{equation}

What is the precise geometric interpretation of the mapping defined by Eq.~(\ref{GARotDef})? Although the bivector $B$ need not be a blade, it can always be written as a sum of commuting bivector blades (consult Ch.~3-4 in \cite{Hestenes}),
\begin{equation}
B = B_1 + \ldots + B_m ,
\end{equation}
where $B_j B_k = B_k B_j$ for $j\neq k$, and the exponential in (\ref{GARotorDef}) can then be factorized,
\begin{equation}
R = R_m \ldots R_1 = e^{-B_m/2} \ldots e^{-B_1/2} .
\end{equation}
Since $\rev{R}=\rev{R}_1 \ldots \rev{R}_m$, the generic rotation (\ref{GARotDef}) is expressible as a commuting composition of simple rotations,
\begin{equation}
a' = R_m (\ldots (R_1 a \rev{R}_1) \ldots) \rev{R}_m .
\end{equation}
Each simple rotation $R_j a \rev{R}_j = e^{-B_j/2} \,a\, e^{B_j/2}$ has a direct geometric interpretation. It rotates vectors parallel to $B_j$ by an angle $|B_j|$ in the direction dictated by the orientation of $B_j$, while leaving the vectors perpendicular to $B_j$ unchanged.

Let us now consider a situation where $A$ is a blade (of grade $\geq 2$), and $B \in \mathcal{G}(A)$ is a generic bivector. Since $B A = B\cdot A = A \cdot B = A B$, we observe that 
\begin{equation} \label{GARotSubspace}
R A \rev{R} = R \rev{R} A = A .
\end{equation} 
As a consequence, $A$ represents an invariant subspace of the rotation characterized by $B$, meaning that the image of any (multi)vector from $\mathcal{G}(A)$ is again a (multi)vector in $\mathcal{G}(A)$. On the other hand, vectors perpendicular to $A$ are not rotated at all, as can be seen from the expansion (\ref{GARotExpansion}), and the fact that $a\cdot B=0$.

%%%%%%%%%%%%%%%%%%%%%%%%%%%%%%%%%%%%%%
\subsection{Transformations and induced mappings} \label{sec:TrGC}
%%%%%%%%%%%%%%%%%%%%%%%%%%%%%%%%%%%%%%

Let $f$ be an invertible smooth mapping (a diffeomorphism) on the space of partial observables $\mathcal{C}$ that maps points $q$ to $q' = f(q)$.  For a vector $a$ in the tangent space of $\mathcal{C}$, the derivative of $f$ in direction $a$ is defined
\begin{equation} \label{GAdifMap}
\underline{f}(a;q) 
\equiv a \cdot \partial_q f(q)
:= \lim_{\eps \rightarrow 0} \frac{f(q+\eps a) - f(q)}{\eps} ,
\end{equation}
and gives rise to a $q$-dependent linear function, the \emph{differential} of $f$, that maps vectors $a$ at point $q$ to vectors $\underline{f}(a)$ at $q'$. It is natural to extend the domain of $\underline{f}$ to general multivectors by demanding linearity and the \emph{outermorphism} property \cite[Ch. 3-1]{Hestenes}
\begin{equation}
\underline{f}(A \wedge B) = \underline{f}(A) \wedge \underline{f}(B) .
\end{equation}

The \emph{adjoint} of $\uline{f}$, denoted $\oline{f}$, is defined
\begin{equation} \label{GCadjoint}
\overline{f}(b;q) := \partial_q f(q) \cdot b ,
\end{equation}
so that for any two vectors $a$ and $b$, the identity 
\begin{equation}
b \cdot \underline{f}(a) = \overline{f}(b) \cdot a
\end{equation}
holds. It maps vectors $b$ at a point $q'$ to vectors $\oline{f}(b)$ at $q$, and extends to an outermorphism in the same way as $\uline{f}$.
Let us add, for completeness, that for scalar arguments, we define $\underline{f}(\alpha) = \overline{f}(\alpha) = \alpha$. We will refer to the differential and the adjoint collectively as the induced mappings.

For an $r$-vector $A_r$ and an $s$-vector $B_s$, the following useful identities hold \cite[Ch. 3-1]{Hestenes}
\begin{align} \label{GAadjointIdent}
A_r \cdot \overline{f}(B_s) = \overline{f}[ \underline{f}(A_r) \cdot B_s ] ~~~~~ {\rm for}~ r \leq s , \nonumber\\
\underline{f}(A_r) \cdot B_s = \underline{f}[ A_r \cdot \overline{f}(B_s) ] ~~~~~ {\rm for}~ r \geq s .
\end{align}

Let us consider an arbitrary multivector-valued function $F$ on $\mathcal{C}$. The chain rule for differentiation,
\begin{align} \label{GCchainRule}
a \cdot \partial_q F(f(q)) 
&= \lim_{\eps \rightarrow 0} \frac{F \big(f(q+\eps a) \big) - F(f(q))}{\eps} \nonumber\\
&= \lim_{\eps \rightarrow 0} \frac{F \big(f(q)+\eps \uline{f}(a) \big) - F(f(q))}{\eps} \nonumber\\
&= \uline{f}(a) \cdot \partial_{q'} F(q') ,
\end{align}
shows that the vector derivative operator transforms under $f$ as
\begin{equation} \label{GCTrDeriv}
\partial_q = \oline{f}(\partial_{q'}) .
\end{equation}
Now, since $f$ is invertible, we may choose in Eq.~(\ref{GCchainRule}) $F = f^{-1}$ to obtain
\begin{equation}
\uline{f^{-1}} \, \uline{f}(a) = a .
\end{equation}
From the latter equation we deduce
\begin{equation}
\uline{f^{-1}} = \uline{f}^{-1} ~~~{\rm and}~~~
\oline{f^{-1}} = \oline{f}^{-1} ,
\end{equation}
which enables us to choose the way of writing inverse induced mappings. 

Under $f$, vectors are naturally transformed by the induced differential, $a'=\uline{f}(a)$. It is also possible to consider another transformation rule, obeyed by \emph{covectors}, $b'=\oline{f^{-1}}(b)$. The latter definition ensures that the inner product between a vector and a covector is preserved under the action of $f$,
\begin{equation}
a' \cdot b' = \uline{f}(a) \cdot \oline{f^{-1}}(b) = \uline{f}^{-1} \uline{f} (a) \cdot b = a \cdot b .
\end{equation}
A typical representative of covectors is the vector derivative operator $\partial_q$ (see Eq.~(\ref{GCTrDeriv})).

The same distinction can be made for arbitrary multivectors. For example, the infinitesimal surface element transforms as a (multi)vector, $d\Gamma'=\uline{f}(d\Gamma)$, as follows from the law of integral substitution. On the other hand, we have seen in Sec.~\ref{sec:Sym} that the momentum transforms naturally as a (multi)covector, $P'=\oline{f^{-1}}(P)$.

The induced transforms $\uline{f}$ and $\oline{f}$ are functions of $q$ which can be further differentiated. This results in second derivatives of $f$. The geometric algebra notation allows to express the symmetry of the second directional derivatives of $f$, usually stated for arbitrary $q$-independent vectors $a$, $b$ and $c$ in the form
\begin{equation}
0 = a\cdot\partial_q b\cdot\partial_q f\cdot c - b\cdot\partial_q a\cdot\partial_q f \cdot c
= (b \wedge a) \cdot (\partial_q \wedge \partial_q ) f(q) \cdot c ,
\end{equation}
in a neat equivalent way
\begin{equation}
\partial_q \wedge \oline{f}(c) = 0 .
\end{equation}
This formula extends by the Leibniz product rule,
\begin{equation}
\partial_q \wedge \oline{f}(c_1 \wedge c_2)
= \partial_q \wedge \oline{f}(c_1) \wedge \oline{f}(c_2)
= \dot{\partial}_q \wedge \dot{\oline{f}}(c_1) \wedge \oline{f}(c_2) + \dot{\partial}_q \wedge \oline{f}(c_1) \wedge \dot{\oline{f}}(c_2) = 0 ,
\end{equation} 
even if the vector $c$ is replaced by an arbitrary constant multivector $A$, 
\begin{equation} \label{GCadjointCurl}
\partial_q \wedge \oline{f}(A) = 0 .
\end{equation}
(Overdots have been used to control the scope of differentiation.)

Let us now specialize to infinitesimal diffeomorphisms
\begin{equation} \label{GCinsfmDiffeo}
f(q) = q + \eps v(q) .
\end{equation}
The action of the differential on an arbitrary $r$-blade $A_r = \blade{a}{r}$ is given by
\begin{equation}
\uline{f}(A_r) 
= (a_1 + \eps \, a_1 \cdot \partial_q v) \wedge \ldots \wedge (a_r + \eps \, a_r \cdot \partial_q v)
= A_r + \eps (A_r \cdot \partial_q) \wedge v + O(\eps^2) .
\end{equation}
For the adjoint outermorphism, we then find
\begin{equation}
\oline{f}(B_r) \cdot A_r
= B_r \cdot \uline{f}(A_r) 
= B_r \cdot A_r + \eps \big( \dot{\partial}_q \wedge (\dot{v} \cdot B_r) \big) \cdot A_r + O(\eps^2) .
\end{equation} 
By the linearity of the above expressions we therefore conclude that for an arbitrary multivector $A$,
\begin{align} \label{GCinfsmTr}
\uline{f}(A) &\approx A + \eps (A \cdot \partial_q) \wedge v , \nonumber\\
\oline{f}(A) &\approx A + \eps \, \dot{\partial}_q \wedge (\dot{v} \cdot A) ,
\end{align}
up to the first order in $\eps$. In this approximation, the inverse of $f$ reads 
\begin{equation}
f^{-1}(q) = q - \eps v(q) ,
\end{equation}
and so we immediately obtain, in addition,
\begin{align} \label{GCinfsmInvTr}
\uline{f}^{-1}(A) &\approx A - \eps (A \cdot \partial_q) \wedge v , \nonumber\\
\oline{f^{-1}}(A) &\approx A - \eps \, \dot{\partial}_q \wedge (\dot{v} \cdot A) .
\end{align}

At the end of this section we briefly consider an example of a mapping between two distinct manifolds, which we use in our treatment of the scalar field theory in Sec.~\ref{sec:ScF}. There, we represent the motions $\gamma$ by a mapping
\begin{equation}
g(x) = x + y(x)
\end{equation}
from the spacetime (a linear $D$-dimensional subspace of $\mathcal{C}$ with a pseudoscalar $I_x$) to $\mathcal{C}$. 
Any spacetime blade $A_r=\blade{a}{r} \in \mathcal{G}(I_x)$ is mapped by the associated outermorphism 
\begin{equation}
\underline{g}(a;x) = a \cdot \partial_x g(x) = a + a \cdot \partial_x y(x)
\end{equation}
to a blade $\underline{g}(A_r)$ in the tangent algebra of $\gamma$, $\mathcal{G}(d\Gamma)$. Here, $\partial_x$ denotes the vector derivative with respect to a spacetime point $x$. 

Concretely, we find the expansion
\begin{equation} \label{GCgdifmap}
\underline{g}(A_r) 
= (a_1+a_1\cdot\partial_x y) \wedge \ldots \wedge (a_r+a_r\cdot\partial_x y) 
= A_r + (A_r \cdot \partial_x ) \wedge y + \ldots ,
\end{equation}
where the ellipsis gathers terms with two or more $y$'s. For example,
\begin{equation} \label{GCgdGamma}
d\Gamma = \uline{g}(dX) 
= dX + (dX \cdot \partial_x ) \wedge y + \ldots ,
\end{equation}
and (recall the identities (\ref{GAadjointIdent}))
\begin{equation} \label{GCgdGammaPartial}
d\Gamma \cdot \partial_q 
= \uline{g}(dX) \cdot \oline{g^{-1}}(\partial_x)
= \uline{g}(dX \cdot \partial_x)
= dX \cdot \partial_x + \big( (dX \cdot \partial_x) \cdot \dot{\partial}_x \big) \wedge \dot{y} + \ldots ,
\end{equation}
where $dX=|dX|I_x$. ($|dX|$ is the unoriented spacetime measure, commonly denoted by $d^Dx$.) 

The differential operator $dX \cdot \partial_x$ acts on all functions that may appear to its right. Whether these include also $y(x)$ in the second term on the right-hand side of Eq.~(\ref{GCgdGammaPartial}) has no effect, since $(dX \cdot \partial_x) \cdot \dot{\partial}_x = dX \cdot (\partial_x \wedge \dot{\partial}_x)$, and $\partial_x \wedge \partial_x \, y(x) = 0$.

%%%%%%%%%%%%%%%%%%%%%%%%%%%%%%%%%%%%%%
\section{Transformation of canonical equations of motion} \label{sec:TrPhysMotCanEOM}
%%%%%%%%%%%%%%%%%%%%%%%%%%%%%%%%%%%%%%

We will show that if
\begin{equation}
f: q \mapsto q'=f(q)
\end{equation}
is a generic diffeomorphism on the configuration space $\mathcal{C}$, and $\gamma$, $P$ and $\lambda$ satisfy the canonical equations of motion (\ref{CanEOM}), then $\gamma'$, $P'$ and $\lambda'$, defined by
\begin{align}
\gamma' &= \{q'=f(q)\,|\,q \in \gamma \} \\
P'(q') &= \overline{f^{-1}}(P(q);q) \\
\lambda'(q') &= \lambda(q) ,
\end{align}
satisfy the canonical equations of motion with a new Hamiltonian
\begin{equation} \label{DefHPrime}
H'(q',P') = H(q,P) .
\end{equation}

Let us recall that the surface element of $\gamma$, and the vector derivative $\partial_q$ transform under $f$ as 
\begin{equation}
d\Gamma'(q') = \underline{f}(d\Gamma(q);q) \quad,\quad
\partial_{q'} = \overline{f^{-1}}(\partial_q) .
\end{equation}

The proof for the first canonical equation (\ref{CanEOM1}) proceeds as follows:
\begin{equation} \label{CanEOM1Primed}
\lambda' \partial_{P'} H'(q',P') 
= \lambda \underline{f}(\partial_P) H'(q',\overline{f^{-1}}(P))
= \underline{f} \big( \lambda\, \partial_P H(q,P) \big)
= \underline{f}(d\Gamma)
= d\Gamma' .
\end{equation}

The proof for the second canonical equation (\ref{CanEOM2}) is more involved. Starting with the right-hand side for $D>1$ (for $D=1$ the proof is fully analogous), 
\begin{equation} \label{TrEOM2}
(d\Gamma' \cdot \partial_{q'} ) \cdot P'
= \big( \underline{f}(d\Gamma) \cdot \overline{f^{-1}}(\partial_q) \big) \cdot \overline{f^{-1}}(P)
= \overline{f^{-1}} \big( (d\Gamma \cdot \partial_q) \cdot P \big) + \big( \underline{f}(d\Gamma) \cdot \overline{f^{-1}}(\dot{\partial}_q) \big) \cdot \dot{\overline{f^{-1}}} (P) ,
\end{equation}
where we have repeatedly used Formulas~(\ref{GAadjointIdent}).
To compare with the left-hand side of Eq.~(\ref{CanEOM2}), we differentiate
\begin{equation}
\dot{\partial}_q H(\dot{q},P)
= \dot{\partial}_q H'(\dot{f}(q),\dot{\overline{f^{-1}}}(P))
= \overline{f}(\dot{\partial}_{q'}) H'(\dot{q'},P') 
+ \dot{\partial}_q \dot{\overline{f^{-1}}}(P) \cdot \partial_{P'} H'(q',P') ,
\end{equation}
and substitute this result into the first term in Eq.~(\ref{TrEOM2}) using the ``unprimed" Eq.~(\ref{CanEOM2}). With Eq.~(\ref{CanEOM1Primed}) replacing the $P'$-differentiation of $H'$ by $d\Gamma'$, we arrive at
\begin{equation}
(d\Gamma' \cdot \partial_{q'} ) \cdot P'
= (-1)^D \lambda' \dot{\partial}_{q'} H'(\dot{q'},P') 
+ (-1)^D \overline{f^{-1}} (\dot{\partial}_q) \dot{\overline{f^{-1}}}(P) \cdot \underline{f}(d\Gamma)
+\big( \underline{f}(d\Gamma) \cdot \overline{f^{-1}}(\dot{\partial}_q) \big) \cdot \dot{\overline{f^{-1}}} (P) .
\end{equation}

Now, the second and the third term can be combined by means of the identity (\ref{GAident1}) to yield
\begin{equation}
\underline{f}(d\Gamma) \cdot \big( \overline{f^{-1}}(\dot{\partial}_q) \wedge \dot{\overline{f^{-1}}} (P)  \big) .
\end{equation} 
But this expression vanishes, since, by Eq.~(\ref{GCadjointCurl}),
\begin{equation}
\dot{\partial}_q \wedge \overline{f} \, \dot{\overline{f^{-1}}} (P)
= - \dot{\partial}_q \wedge \dot{\overline{f}} \,\overline{f^{-1}} (P)
= 0 .
\end{equation}
Therefore, we find the desired result
\begin{equation}
(d\Gamma' \cdot \partial_{q'} ) \cdot P'
= (-1)^D \lambda' \dot{\partial}_{q'} H'(\dot{q'},P') .
\end{equation}

Finally, the third canonical equation (\ref{CanEOM3}) is fulfilled trivially by definition (\ref{DefHPrime}).

%%%%%%%%%%%%%%%%%%%%%%%%%%%%%%%%%%%%%%

\end{document}